\documentclass{aa}  
\usepackage{graphicx}
\usepackage{txfonts}
\def\be{\begin{equation}}
\def\ee{\end{equation}}

\def\be{\begin{equation}}
\def\ee{\end{equation}}
\catcode`\@=11 
\def\@versim#1#2{\vcenter{\offinterlineskip
\ialign{$\m@th#1\hfil##\hfil$\crcr#2\crcr\sim\crcr } }}
\def\lsim{\mathrel{\mathpalette\@versim<}}
\def\gsim{\mathrel{\mathpalette\@versim>}}

\begin{document}
   \title{On the Radio/X-ray Correlation in Microquasars}

   \author{Y. Q. Xue
          \inst{1}
	  \and
	  W. Cui\inst{1}
          }

   \offprints{Y. Q. Xue}

   \institute{Department of Physics, Purdue University, West Lafayette, IN 47907\\ \email{xuey@physics.purdue.edu, cui@physics.purdue.edu}
             }

   \date{Received 23 Nov. 2006 / Accepted 08 Feb. 2007}

 
  \abstract
  {The broadband spectral energy distribution of microquasars has proven to be a valuable 
  tool for assessing the roles of jets and accretion flows in microquasars, as well as the 
  coupling between the two. The coupling might manifest itself observationally in the 
  correlated radio and X-ray variabilities of a source. Such a radio/X-ray correlation has
  indeed been seen in several microquasars during the low-hard state and subsequently been 
  claimed to be universal for all. If proven, the universal correlation would have profound 
  implications on theoretical models. However, there is already observational evidence that 
  suggests otherwise.}
  {In this paper, we critically examine the radio/X-ray correlation in a sample of microquasars,
  in the low-hard state as well as during state transitions, with a goal of testing the claimed 
  universality of the correlation on observational grounds.}
  {We have assembled a comprehensive data set from the simultaneous/contemporaneous radio and 
  X-ray observations of representative microquasars. The data have allowed us to quantify 
  the radio/X-ray correlation on a source by source basis. }   
  {We find that the radio/X-ray correlation of microquasars exhibits diverse behaviors, 
  both in the low-hard and transitional states, ranging from being very week (or none at 
  all) to very strong. There is even a hint of spectral dependence of the correlation 
  in some cases. }
  {Our results rule out the claimed universality of the radio/X-ray correlation of 
  microquasars, even for the low-hard state. On the other hand, we do find that the radio
  and X-ray variabilities are, to varying degrees, correlated in most cases. }


\titlerunning{Radio/X-ray Correlation in Microquasars}
\authorrunning{Y. Q. Xue \& W. Cui}

   \maketitle
%

\section{Introduction}

Microquasars are inherently broadband objects, like active galactic nuclei and gamma-ray
bursts. Roughly speaking, the spectral energy distribution (SED) of microquasars can be
divided into three main regimes (see review by, e.g., Liang 1998): a power-law distribution
at radio frequencies, which could extend up to infrared/optical regions (e.g., Chaty et 
al. 2003), a blackbody-like profile in the middle (from optical to soft X-ray frequencies), 
and a power-law distribution again at hard X-ray/soft gamma-ray energies (which eventually 
rolls over). To interpret the SED of microquasars, two classes of models have been 
proposed, with one being accretion-based (e.g., Esin et al. 2001) and the other jet-based 
(Markoff et al. 2001, 2003). However, neither class of models can satisfactorily explain 
the {\it overall} SED of microquasars. The accretion-based models typically fail at low 
frequencies (radio and infrared in particular; see discussion in Yuan, Cui, \& Narayan 
2005), while the jet-based models suffer from a number of issues (Zdziarski et al. 2003). 
A hybrid scenario seems more likely (and natural), in which both the accretion flows and 
jets make critical contributions to the SED of microquasars but the relative importance of 
the contributions may vary at different frequencies. For example, the coupled accretion-jet 
model proposed by Yuan et al. (2005) postulates that the radio (and a bulk of infrared) 
emission comes entirely from the jets and the X-ray (and soft gamma-ray) emission is due 
almost entirely to the accretion flows. An effective way to assess the roles of the jets 
and accretion flows in microquasars is, therefore, to study the broadband SED of these 
systems. 

A key ingredient in nearly all models on jet formation is the physical coupling between
the jets and accretion flows (e.g., Falcke \& Biermann 1995; Meier 2001). For microquasars, 
such a jet-accretion coupling might manifest itself in the correlated radio and X-ray 
variabilities (Robertson \& Leiter 2004; Yuan \& Cui 2005). Indeed, a positive radio/X-ray 
correlation was first
established for GX 339--4 in the low-hard state (Corbel et al. 2003). A similar correlation 
was subsequently also seen in V404 Cyg (also in the low-hard state), which led to the 
claim that the correlation is the same for all microquasars (Gallo et al. 2003). If proven, 
such a universal radio/X-ray correlation would have profound implications on theoretical 
models (e.g., Heinz 2004; Robertson \& Leiter 2004; Yuan \& Cui 2005). However, as noted by 
Xue et al. (2006), the data shown by Gallo et al. (2003) are dominated 
by only two sources (GX 339--4 and V404 Cyg); the data points for other sources are very 
limited in their dynamical ranges yet already show significant deviations from the universal 
correlation. Subsequent studies provided additional evidence for such deviations (see, e.g., 
Fig.~5 in Choudhury et al. 2003 and Fig.~17 in Wilms et al. 2006), even for the low-hard 
state, although the authors tended to be dismissive of the discrepancies. Based on these and 
other observations, Xue et al. (2006) suggested that, while the radio and X-ray variabilities 
could well be correlated among microquasars individually, the correlation might not be 
quantitatively the same for all.

To resolve this important issue, we have undertaken an effort to systematically examine
the simultaneous radio and X-ray data already taken for a representative sample of 
microquasars. We believe that the best way to separate personality from universality is 
to carefully examine the radio/X-ray correlation on a source by source basis. We have
also taken the opportunity to examine the radio/X-ray correlation beyond the low-hard 
state, to the extent that the data allow.

\section{Data}

We first carried out a comprehensive search for radio observations of microquasars in the 
archival databases and in the literatures. For each radio observation found, we then 
checked the availability of simultaneous/contemporaneous X-ray data on the same source. 
For this work, we used only data sets in which the corresponding radio and X-ray 
observations were taken within one day of each other. In the end, we found over 300 
sets of the radio/X-ray observations that satisfy our criteria. Of the total, about 25\% 
contain the radio and X-ray observations taken simultaneously, about 50\% observations 
taken within 5 hours of each other, and about 88\% observations taken within 12 hours of 
each other.

\subsection{Radio Data}

Table 1 summarizes the sources of the radio data used in this work, as well as the year(s)
in which the data were taken for each microquasar in the sample. The data come from 
observations made with a number of radio observatories, including the Very Large Array, 
the Australia Telescope Compact Array, the Molonglo Observatory Synthesis Telescope, the Ryle 
Telescope, the Multi-Element Radio-Linked Interferometer Network, the Giant Metrewave Radio 
Telescope, and the Green Bank Interferometer, and, in most cases, at more than one 
frequency (see the listed references for details on the observations).

\subsection{X-ray Data}

The X-ray data were derived exclusively from the {\it Rossi X-ray Timing Explorer} 
({\it RXTE}) archival database. For this work, we used only data from the Proportional 
Counter Array (PCA). The PCA consists of five nearly identical proportional counter 
units (PCUs) but not all of the PCUs are always turned on for a given observation. 
For the observations of interest here, PCU 2 was always in operation. For simplicity, 
we used only the PCU 2 data, which already provide sufficient statistics for our
purposes.

The {\it RXTE} data were reduced with {\em FTOOLS 6.0.1} (Blackburn 1995), following the
usual procedure that we have adopted (see, e.g., Cui 2004; Xue \& Cui 2005). For each 
observation, we first filtered data by following the standard procedure for bright 
sources, which resulted in a list of good time intervals (GTIs). We then simulated
background events for the observation with the latest background model
that is appropriate for bright sources (pca\_bkgd\_cmbrightvle\_eMv20051128.mdl).\footnote{The only exception is XTE J1118+480 
in its 2005 outburst, in which the source was quite faint and we had to use the
background model for faint sources (pca\_bkgd\_cmfaintl7\_eMv20051128.mdl).} Using the 
GTIs, we then produced the PCU 2 spectrum from the {\it Standard2} 
data for the observation. Note that we used only data from the first xenon layer, which 
is best calibrated. We repeated the steps to derive a corresponding background spectrum 
for PCU 2 from the simulated background events.

We carried out the spectral modeling in XSPEC 12.2.0ba (Arnaud 1996). Because the {\it RXTE}
data have poor signal-to-noise ratios at low energies, during the spectral fitting we 
fixed the hydrogen column density for each source at the Galactic value along the line 
of sight (Dickey \& Lockman 1990). Also, because the main purposes of the spectral modeling 
here is to quantify the shape of the SED and derive accurate X-ray fluxes, we limited 
ourselves to simple empirical models, including power law and power law plus multi-color 
disk (``diskbb'' in XSPEC), with or without a high-energy rollover (``highecut'' in XSPEC). 
In some cases, we also needed an additional Gaussian component and sometimes an absorption 
edge (``smedge'' in XSPEC) to account for the residuals. A discussion of the physical
nature of each component in the adopted models is beyond the scope of this work. We 
limited the spectral fits to an energy range of 3--30 keV (which is nominally covered by the
first xenon layer) and added 1\% systematic uncertainty to the data. Statistically 
acceptable fits (i.e., with the reduced $\chi^2$ values around unity) were obtained for 
all observations. In each case, the SED was then derived by using the best-fit model to 
unfold and de-absorb the measured spectrum.

\section{Results}

To facilitate direct comparisons with the published results (e.g., in Gallo et al. 2003), 
we adopted the energy band of 2--11 keV for computing X-ray fluxes. The X-ray fluxes 
presented in this work have all been corrected for interstellar absorption.

\subsection{Diversity of the Radio/X-ray Correlation in Microquasars}

Figure~1 summarizes the diverse behaviors of the radio/X-ray correlation that we found 
among individual microquasars. In the figure, the radio flux density ($F_R$) and the 
X-ray flux ($F_X$) are shown for four selected microquasars in our sample that are all in 
the low-hard state.

Cyg X-1 is representative of persistent microquasars (which are in the minority of the
population). It is perhaps still the most frequently observed and studied member of the 
class. A tremendous amount of X-ray data have been collected on Cyg X-1. For this work,
we selected data from the {\it RXTE} observations of the source in 1999, because of the
large dynamical range of the radio data (Gleissener et al. 2004). The results show that,
while there seems to be a general positive correlation between the radio and X-ray 
variabilities of the source, the correlation is a fairly loose one. This is in general 
agreement with the conclusion of Wilms et al. (2006) based on some of the same data (see 
their Fig.~17). Fitting the data with a power law we found $F_R \propto F_X^{0.5 \pm 0.4}$.
Obviously, the fit is very poor, as shown in Fig.~1(1). For comparison, Wilms et al. 
found $F_R \propto F_X^{1.05}$ when they included additonal low-hard state data in their 
analysis. They speculated that the observed deviation from the universal radio/X-ray 
correlation might be related to the fact that Cyg X-1 is a persistent source. However, if the 
radio/X-ray correlation is related to the coupling of the jets and accretion flows, 
it would not be obvious why it has anything to do with a source being persistent or 
transient.

Together with GRS~1915+105, GRO~J1655--40 is among the first microquasars whose jets are
seen to exhibit superluminal motions (Mirabel \& Rodriguez 1999). In fact, the study of 
these sources led to the use of the term ``microquasar'' in the literatures, highlighting 
their similarities to supermassive black hole systems. GRO~J1655--40 is representative of
transient microquasars (which is the dominant majority). After spending many years in 
quiescence, it underwent a major outburst in 2005, during which it became the target of 
intensive monitoring efforts both in the radio and X-ray wavebands. Some of the results 
based on those observations have been reported by Shaposhnikov et al. (2006), including 
the radio/X-ray correlation. We re-analyzed the data here for two reasons. One is to 
compare directly with other sources in the sample, highlighting the diverse behaviors, 
and the other is to examine the possible spectral-dependence of the radio/X-ray correlation 
(see the next section). Confirming the results of Shaposhnikov et al. (2006), Fig~1(2) 
shows that the source varied greatly in X-rays in the low-hard state but hardly varied at 
all in radio. Fitting the data with a power law, we found $F_R \propto F_X^{-0.11 \pm 0.04}$. 
Note that the fit is not statistically acceptable in this case either. This result 
shows the most dramatic departure from the claimed universal radio/X-ray correlation. 

XTE J1118+480 was the focus of multiple intensive multi-wavelength campaigns during its
outburst in 2000 (e.g., Hynes et al. 2000; McClintock et al. 2001; Frontera et al. 2001; 
Chaty et al. 2003). Intriguingly, it remained in the low-hard state throughout the 
outburst, which makes it representative of a small group of transient microquasars 
(including, e.g., V404 Cyg and GRO~J0422+32) that are known to do the same. Unfortunately,
the dynamical range of the simultaneous radio and X-ray observations is
too limited during the outburst, which
is hardly adequate for investigating radio/X-ray correlations. Luckily, the source was
seen in outburst again in 2005. It was intensely followed during this outburst both in 
the radio and X-rays. Fig.~1(3) shows the 2005 data. Fitting the
data with a power law, we found $F_R \propto F_X^{0.71 \pm 0.03}$ and the fit is excellent. 
In other words, XTE~J1118+480 happens to nicely follow the claimed universal radio/X-ray 
correlation over a flux range that spans roughly two orders of magnitude both in radio and 
X-rays. In fact, this is the tightest radio/X-ray correlation that has been seen among all 
microquasars.

V4641~Sgr is a member of the ever-growing population of fast X-ray transients. It was
discovered during a giant outburst in 1999 that lasted only for a few days! During the
outburst, the jets with superluminal motion were seen. The unusual characteristics of the
source generated a lot of interests and consequently made it the target of a number of 
observational efforts. Unfortunately, there were no simultaneous/contemporaneous 
radio/X-ray observations during the most active period of the source, partly due to the
briefness of the outburst. Fig.~1(4) summarizes the results of all radio/X-ray measurements 
taken at nearly the end of this outburst and during smaller outbursts between 2002 and 2004, 
when the source was in the low-hard state. The radio and X-ray variabilities are again 
positively correlated but, taken at its face value, the correlation is quite complex in 
this case (see, however, discussion in the next section). At the low end of X-ray fluxes, 
the source hardly varied in X-rays while exhibited significant variability in the radio 
band. At higher fluxes, the two bands begin to show a tighter correlation. If we fit only 
the high-flux ``branch'' of the data set (i.e., the seven data points toward the high end 
of the fluxes), we would get $F_R \propto F_X^{1.14 \pm 0.01}$ (as indicated by the dotted 
line in the figure). 

\subsection{Complexity of the Radio/X-ray Correlation in Microquasars}

We proceeded to investigate possible spectral dependence of the observed radio/X-ray 
correlations. We selected sources in the sample for which multi-frequency radio data 
are available. Figure~2 summarizes the results. Like in Fig.~1, only data for the
low-hard state are shown in Fig.~2. 

As already mentioned, GX~339--4 is one of the two sources on which the claimed universal
radio/X-ray correlation is based. The tight correlation is apparent in Fig.~2(1), as
first reported by Corbel et al. (2003), using only the 8.64 GHz data. We note, however,
that we have added a new data point that was taken during a more recent outburst (in 
2002) and that the new data point is clearly below the extrapolation of the reported 
correlation. The phenomenon is also 
evident for XTE~J1118+480, as shown in Fig.~2(2), which now also includes data points 
from the 2000 outburst. There are two scenarios to consider. One is that the radio/X-ray 
correlation in these sources might level off at high fluxes; the other is that the 
normalization (or even the slope) of the correlation might be different for different 
outbursts. Unfortunately, the existing data are not sufficient to allow us to distinguish 
the two scenarios. In either case, the results have added a new twist to the radio/X-ray
correlation in microquasars that any viable model must try to account for. It is also
worth noting that the loose correlation in Cyg X-1 and the complex correlation in
V4641 Sgr (see Fig.~1) might also be the manifestation of the same phenomenon.

Examining the radio/X-ray correlation in each radio band separately, we found that the
4.80 GHz data can be best described as $F_R \propto F_X^{0.86\pm 0.03}$ and the 8.64 GHz 
data as $F_R \propto F_X^{0.71\pm 0.01}$ for GX 339--4, when we excluded the 2002 data
points. This result seems to suggest a spectral dependence of the radio/X-ray correlation,
at least for GX~339--4 in the low-hard state. We repeated the analysis for XTE~J1118+480, 
which also shows a tight power-law correlation. Excluding data from the 2000 outburst,
we found that the 4.86 GHz data can be best described as $F_R \propto F_X^{0.66\pm 0.03}$,
the 8.46 GHz data as $F_R \propto F_X^{0.71\pm 0.03}$, and the 22.46 GHz data as
$F_R \propto F_X^{0.66\pm 0.07}$. Therefore, there is no apparent frequency dependence of 
the slope of the radio/X-ray correlation for XTE~J1118+480, although there appears to be
a shift in the overall normalization among different radio bands in this case. The 
phenomenology is, therefore, complex in this regard as well. For completeness, we also 
show, in Fig.~2, the multi-band radio measurements for GRO~J1655--40 and V4641 Sgr. It is 
not obvious how to best quantify the radio/X-ray correlation in these cases. By eye, no 
frequency dependence of the correlation is apparent in either case.

\subsection{The Radio/X-ray Correlation beyond the Low-Hard State}

The claimed universal radio/X-ray correlation is said to be applicable mainly to the
low-hard state of microquasars (Gallo et al. 2003). Unfortunately, the spectral states are, 
in general, not well defined, as discussed in great detail in Xue et al. (2006). The degree 
of arbitrariness in the definitions has made the discussion and comparison of results very 
difficult. Xue et al. (2006) suggested that the spectral states should be defined simply based on the 
overall shape of the SED. The low-hard state and high-soft state thus defined 
would, unfortunately, be only of theoretical value, because they are nearly impossible to 
catch in practice. These two diametrically-opposed states represent extreme scenarios in 
which only cold or hot accretion flows contribute to the X-ray emission from microquasars. 
So they might not even occur in nature. In this context, one can only speak of the quasi 
low-hard or high-soft state. We will do the same here. 

Cyg X-1 and GX 339--4 have both shown strong anti-correlations between the radio and X-ray 
variabilities during transitions between the low-hard and high-soft states (e.g., Braes 
\& Miley 1976; Corbel et al. 2000). To see whether this is a generic feature of microquasars, 
we carefully examined the data for the sources in our sample. Figure~3 summarizes the results 
for the transitional states, together with those for the low-hard state for
comparison. We should emphasize that there is hardly any confusion between the low-hard
state and transitional state in the definitions suggested by Xue et al. (2006), which we adopted for this work. 
For clarity, we included only the radio data taken at $\sim$ 4.8 GHz. There are two things worth 
noting. First of all, the flux ranges of the two states overlap, so the flux alone is not a 
reliable predictor of spectral states. This is not new (see, e.g., Homan et al. 2001) and is 
perhaps related to the spectral hysteresis associated with state transitions (e.g., Miyamoto 
et al. 1995; Zdziarski et al. 2004; Xue et al. 2006). Secondly, although the expected anti-correlation 
can be seen in GRO J1655--40 in the transitional state, the radio and X-ray variabilities seem 
to be positively correlated for other sources, including GX 339--4, during transitions, which 
is very puzzling. We caution, however, against drawing any definitive conclusions, given the
sparseness and the very limited dynamical ranges of the data.

Further complication arises when we examine the behavior of XTE~J1859+226 in the transitional
state, as shown in Fig.~4. The ``loop'' pattern shows the lack of one-to-one correspondence 
between the X-ray and radio fluxes, which seems to indicate the presence of some kind of 
hysteresis associated with the variabilities of the source between the X-ray and radio bands.
The phenomenon persists across all radio bands. The data were derived
from observations of XTE J1859+226 during its 1999 outburst. The outburst follows a 
fast-rise-slow-decay profile. Looking at the data more carefully, we found that the 
``loop'' did not vanish after we excluded data from the rising phase of the outburst. 
Therefore, the phenomenon is not related to any hysteresis that might be associated 
with the rise and fall of the source during the outburst.  

\section{Discussion}

We have critically examined the radio/X-ray correlation for microquasars. We have clearly 
demonstrated that there is not a universal relationship that can account for the correlated 
variabilities of microquasars between the radio and X-ray bands, even for the low-hard 
state. In fact, the observed radio/X-ray correlation cannot be adequately described by a 
simple power law in most cases that we have examined. Our results are, therefore, at odds 
with those of Gallo et al. (2003). We believe that the results are quite robust, given
that they are based on a sample that includes representatives of various sub-populations 
of microquasars. On the other hand, there is, to varying degrees, a general positive 
correlation between the radio and X-ray variabilities among the sources in our sample. In
one case (XTE~J1118+480), the correlation is very tight and follows nicely the claimed
universal correlation. Such diverse behaviors present a severe challenge to our 
understanding of the phenomenon.

Besides the lack of universality of the radio/X-ray correlation in microquasars, we also
found further complications, some of which might be generic and some might reflect
the personality of individual sources. Specifically, there seems to be some spectral
dependence of the radio/X-ray correlation, at least for GX~339--4. Therefore, even for a
given source, one might need to specify the spectral bands of interest when one speaks
of the correlation quantitatively. On the other hand, no such spectral dependence is
apparent for XTE~J1118+480, which shows an equally tight (if not tighter) power-law
correlation between the radio and X-ray bands. So, this is probably not a general
phenomenon for all microquasars. Another feature that is worth noting is the possibility 
that the radio/X-ray correlation might exhibit quantitative difference between different
outbursts of the same source. However, the existing data do not allow a definitive
conclusion on the issue.

Examining the data from observations of microquasars in the transitional state, we seem to
confirm the presence of anti-correlation between the radio and X-ray fluxes for
GRO~J1655--40, which has been seen previously. However, for other sources, the radio/X-ray
correlation appears to be positive, qualitatively similar to the low-hard state, which is
puzzling. However, the dynamical range of the data is limited in all cases, so these
results should be taken with a grain of salt. It has been speculated that for microquasars
the jets might be steady in the low-hard state but consist of discrete ejection events 
during the transitions (e.g., Corbel et al. 2000). If this is the case, it would not be
surprising that the radio/X-ray correlation would be, at least quantitatively, different
between the low-hard state and the transitional state. From a theoretical point of view, 
even if the correlation reflects a physical coupling between the jets and accretion flows,
the details of the coupling would not remain the same due, e.g., to the changes in the
accretion flows during state transitions. More data are clearly needed to reveal and
ultimately quantify the difference in the correlation between the states.

Finally, it is puzzling that the $log F_R$--$log F_X$ plot for XTE J1859+226 in the
transitional state shows a loop pattern. This certainly adds another level of complication
to the discussion of the radio/X-ray correlation in microquasars. The phenomenon might 
be an indication of some kind of hysteresis but the hysteresis cannot be associated
with the rising and decaying phases of outbursts. It remains to be seen whether the
phenomenon also occurs in other microquasars.

\begin{acknowledgements}
This work has made use of data obtained through the High Energy Astrophysics Science 
Archive Research Center Online Service, provided by the NASA/Goddard Space Flight 
Center. We gratefully acknowledge financial support from NASA and from the Purdue 
Research Foundation.
\end{acknowledgements}

\begin{table}[ht]
\begin{center}
\caption{Source Sample.}
\begin{tabular}{lcccc}\hline\hline
Source &  Year(s)$^a$ & State(s)$^b$ & Radio Data \\
Name   &  &  & References \\ \hline
Cygnus X-1    &  1999 & LHS & 1 \\
GRO J1655--40 &  2005 & LHS, TS & 2 \\
GX 339--4     &  1997-1999, 2002 & LHS, TS & 3, 4 \\
H 1743--322   &  2003 & LHS, TS & 5-8 \\
V4641 Sgr     &  1999, 2002-2004 & LHS, TS & 9-16 \\
XTE J1118+480 &  2000, 2005 & LHS & 17-19 \\
XTE J1859+226 &  1999 & TS & 20 \\ \hline
\end{tabular}
\end{center}
References: (1) Gleissner et al. 2004;
(2) Brocksopp et al. 2005;
(3) Corbel et al. 2000; (4) Gallo et al. 2004;
(5) Rupen, Mioduszewski, \& Dhawan 2003a; (6) Rupen, Mioduszewski, \& Dhawan 200
3b; (7) Rupen, Mioduszewski, \& Dhawan 2003c; (8) Kalemci et al. 2006;
(9) Ishwara-Chandra \& Pramesh 2005; (10) Uemura et al. 2004; (11) Rupen et al. 
2004; (12) Rupen, Mioduszewski, \& Dhawan 2003d; (13) Rupen, Dhawan, \& 
Mioduszewski 2003; (14) Senkbeil \& Sault 2004; (15) Rupen, Dhawan, \& Mioduszewski 2004
; (16) Hjellming et al. 2000;
(17) Fender et al. 2001; (18) Dhawan et al. 2000; (19) Rupen, Dhawan, \& Miodusz
ewski 2005;
(20) Brocksopp et al. 2002.\\
Note: $^a$ The year(s) in which the data were taken. \\
$^b$ The spectral states covered by the data. LHS stands for ``low-hard state'' and TS 
``transitional state''.
\end{table}

\begin{figure}[ht]
\resizebox{\hsize}{!}{\includegraphics{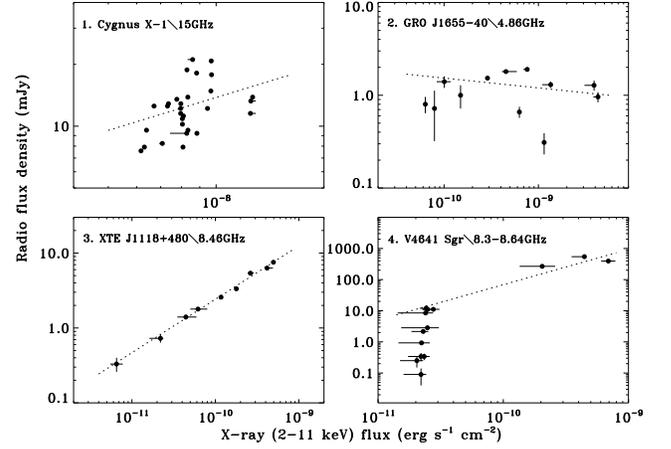}}
\caption{The observed radio/X-ray correlations in the selected microquasars. Only data for the low-hard 
state are shown. The dotted line shows the best power-law fit to the data. Note, however, that
in the case of V4641 Sgr the fit was made only to the seven data points (two of which are too 
close to be seen separately) at high fluxes.}
\end{figure}

\begin{figure}[ht]
\resizebox{\hsize}{!}{\includegraphics{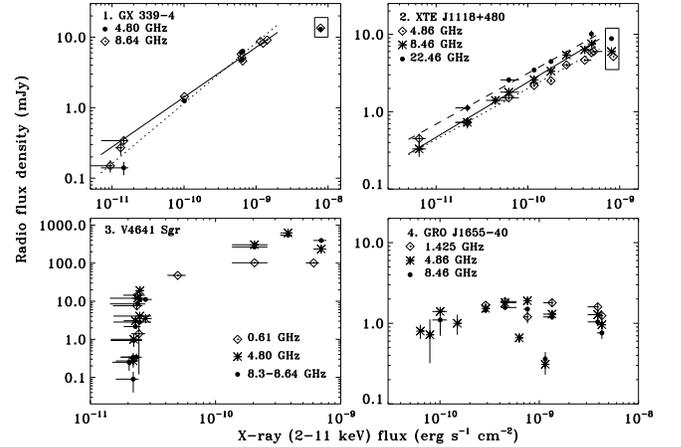}}
\caption{The frequency dependence of radio/X-ray correlations. Radio data in various
bands are shown for the selected sources. The data points in the box were taken during a 
different outburst (see text). Note that only data for the low-hard state are shown. ({\it Panel 1}) The best power-law fit to the 4.80
GHz data is shown in the dotted line and to the 8.64 GHz data in the solid line. ({\it Panel 2}) 
The best power-law fit to the 4.86 GHz data is shown in the dotted line, to the 8.46 GHz data 
in the solid line, and to the 22.46 GHz data in the dashed line. }
\end{figure}

\begin{figure}[ht]
\resizebox{\hsize}{!}{\includegraphics{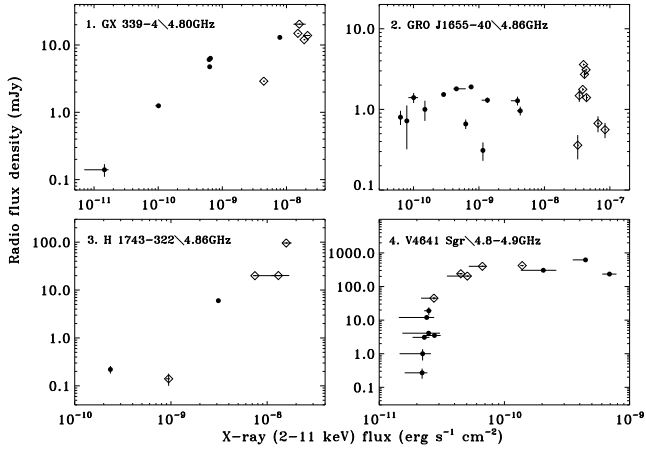}}
\caption{The comparison of radio/X-ray correlations between the low-hard and transitional 
states. The low-hard state data are shown in filled symbols and the transitional state 
data in open symbols. Note that only the 4.8 GHz data are shown here. }
\end{figure}

\begin{figure}[ht]
\resizebox{\hsize}{!}{\includegraphics{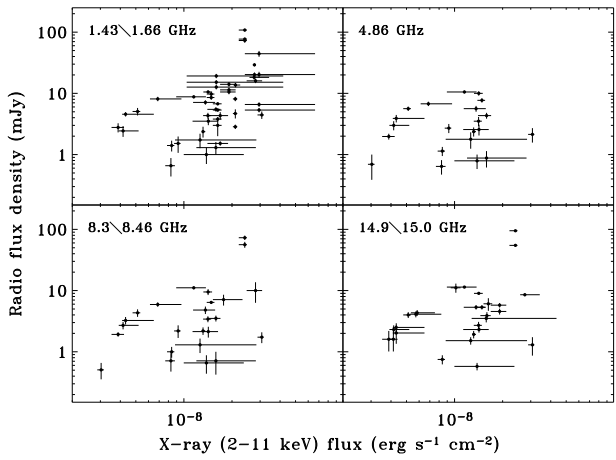}}
\caption{The radio/X-ray correlation for XTE J1859+226 in the transitional state.}
\end{figure}

\end{document}